\documentclass[prb,twocolumn,showpacs,superscriptaddress]{revtex4-1}
\usepackage{graphicx,amsmath,amssymb,amsxtra}
\usepackage{latexsym,psfrag}
\usepackage{graphicx} 

\def\bd{\begin{displaymath}}
\def\ed{\end{displaymath}}
\def\be{\begin{equation}}
\def\ee{\end{equation}}
\def\bea{\begin{eqnarray}}
\def\eea{\end{eqnarray}}
\def\bi{\begin{itemize}}
\def\ei{\end{itemize}}
\def\bn{\begin{enumerate}}
\def\en{\end{enumerate}}

\def\ie{{\it i.e.},\ }

\def\ie{i.e.,\ }

\def\ie{\emph{i.e.,\ }}

\newcommand{\Eq}[1]{Eq.~(\ref{#1})}

\allowdisplaybreaks[1]
\begin{document} 
\title{A minimal model of quantized conductance in interacting ballistic quantum wires} 

\author{Ronny Thomale}
\affiliation{Department of Physics, Princeton University, Princeton, NJ 08544, USA}
\author{Alexander Seidel}
\affiliation  
{Department of Physics and Center for Materials Innovation, Washington University, St. Louis, MO 63136, USA}
\pagestyle{plain}
\begin{abstract}
  We review what we consider to be the minimal model of quantized conductance in a finite
  interacting quantum wire.  Our approach utilizes the simplicity of
  the equation of motion description to both deal with general
  spatially dependent interactions and finite wire geometry. 
  We emphasize the role of two different kinds of boundary conditions,
  one associated with local "chemical" equilibrium in the sense of
  Landauer, the other associated with screening in the proximity of
  the Fermi liquid metallic leads.  The relation of our analysis to
  other approaches to this problem is clarified.  We then use our
  formalism to derive a Drude type expression for the low frequency
  AC-conductance of the finite wire with general interaction profile.
\end{abstract} 
\pacs{72.10.-d, 72.15.Nj, 73.23.-b}

\maketitle

\section{Introduction}
\label{sec:int}
The conductance of one-dimensional quantum wires has been a focus of
research in condensed matter physics for over 20 years.  In agreement
with seminal theoretical works by
Landauer~\cite{landauer57ibm233,landauer70pm863,landauer87zpb} and
B\"uttiker~ \cite{buettiker86prl1761}, early experiments in the late
80's observed a quantization of conductance in units of
$2e^2/h$~\cite{wees-88prl848,wharam-88jpc209}.  This suggested that
there might be a universal explanation for these observations even in
the presence of interactions, which had been neglected in the original
treatment.  However, the precision of measurements of these
conductance plateaus left enough room for the possibility of
non-universal corrections.  Additionally, different theoretical
beliefs concerning the fundamental nature of the conductance did not
converge fast. For nearly one whole decade, the question whether the
quantized conductance is renormalized by interactions in the quantum
wire~\cite{apel-82prb7063,kane-92prl1220,ogata-94prl468} or not~
\cite{kawabata95jpsj30, safi-95prb17040,safi97,ponomarenko95prb8666,
  maslov-95prb5539, alekseev-98prl3503, ponomarenko-99prb16865,safi99ejp451}
dictated the research in this field, where the latter was suggested by
experiments already from the very beginning.
In recent years, the influence and nature of the coupling between the
 leads and the wire on the observed
 value
 of the
conductance has been further emphasized
(e.g. Ref. \onlinecite{imura-02prb035313}, which builds on the notion
of "charge fractionalization" discussed in Ref.
\onlinecite{pham-00prb16397} which is suggested to be observable in
the noise signature of Luttinger liquids~\cite{trauzettel04}).
Moreover, the field had a significant revival when experiments
observed an additional plateau structure generally referred to as 0.7
anomaly~\cite{danneau-08prl016403,thomas-96prl135,yacoby-96prl4612,thomas-00prb13365,kristensen-00prb10950,reilly-02prl246801,biercuk-05prl026801,cronenwett-02prl226805},
and subsequent theoretical studies concentrated on this point as well
as other processes that lead to corrections of the standard quantized
conductance~\cite{meir-02prl196802,matveev04prb245319,sushkov01prb155319,meden06}.
Recently, experimental progress in tunneling spectroscopy of
one-dimensional wire structures promises more detailed studies of the
influence of electron-electron interaction on the conductance and
other transport properties~\cite{chen-09prl036804,auslaender-02s825}.

In this paper, we intend to use an 
approach to the problem which is as elementary as possible.  We show
that the quantized conductance is fully accounted for through the
harmonic equations of motion~\cite{haldane81prl1840} of an
inhomogeneous interacting Luttinger liquid (LL), together with proper
boundary conditions.  We note that the use of equations of motion has
played a role in other works as well.  As a contemporaneous
independent approach to a related work by Safi and
Schulz~\cite{safi-95prb17040}, Maslov and Stone~\cite{maslov-95prb5539}
(MS) have used equation of motion techniques in a rather unusual
geometry, where the Luttinger liquid wire must be assumed infinite as
a matter of principle, and a non-zero electric field is applied only
over a finite region of the infinite wire. This idealized setup is
intimately related to the success of a physically correct but
non-standard application of the Kubo formula, where $\omega$ is taken
to zero before $q$~\cite{maslov-95prb5539} (c.f.  also
Ref.~\onlinecite{safi-95prb17040}).
We argue that the subtleties in
the harmonic fluid approach of MS can be circumvented by imposing
proper boundary conditions at the ends of a finite interacting quantum
wire, which enforce chemical equilibrium between the left lead and the
right-traveling modes of the wire, and vice versa.  Similar boundary
conditions were also emphasized in Refs.~\onlinecite{safi99ejp451}
and~\onlinecite{imura-02prb035313}.
While our physical assumptions are equivalent to those made by Safi
and Schulz~\cite{safi-95prb17040} and MS, it is crucial that we study
such boundary conditions in the presence of general position-dependent
interactions, as will become clear below.  For similar reasons, Safi
and Schulz mainly focus on the case where the Luttinger parameter has
two isolated jumps and is otherwise constant, in a semi-infinite
geometry (this is also the case considered in
Ref. \onlinecite{lebedev-05prb075416}).  There are various
similarities between subsequent work by Safi\cite{safi97,
  safi99ejp451} and the approach developed in the following, and we
will comment on these as well as some important differences as we go
along.


In the following,  we will consider the conductance of
a Luttinger liquid wire with 
interactions that may be arbitrary in strength
(only constrained by the stability of the LL) and profile in the bulk
of the wire.  However, we will assume that they are screened near the
contacts with the metallic leads, as argued in
Ref.~\onlinecite{maslov-95prb5539}.  This introduces a set of boundary
conditions, which we call ``proximity'' conditions.  These are imposed
in addition to the 
boundary condition enforcing "chemical equilibrium" between leads and
chiral modes, to be specified below.
The paper is organized as follows. In Section \ref{sec:ill}, we define
the Luttinger-type Hamiltonian of our model and in particular discuss
our assumptions on proximity boundary conditions. In Section
\ref{sec:emo}, we derive the equations of motion for the current and
review the constraints arising from chemical equilibrium at the
boundary. In Section \ref{sec:dcc} we show that these assumptions
unambiguously imply the quantization of the conductance.  In Section
\ref{sec:acc} we generalize the approach to the case of a low but
finite frequency bias, deriving Drude-type corrections to the DC
result. An explicit formula for the non-universal $\tau$-parameter as
function of spatially dependent interactions is obtained.  In Section
\ref{sec:dis}, we discuss possible modifications of our results for
different experimental setups, and further clarify the relation of our
formalism to other approaches found in the literature.
\section{Inhomogeneous Luttinger Liquid}
\label{sec:ill}
We model the one-dimensional conductor by writing down a Luttinger or
"harmonic fluid" \cite{haldane81prl1840} type of Hamiltonian with
spatially dependent parameters:
\begin{equation}
\begin{split}
H = \frac{1}{\pi}\sum_{\nu}\int dx\, v_{\nu}(x)\{
(\frac{\pi}{2})^2K_{\nu}(x)\Pi_{\upsilon}^{2}(x)\\
+     \frac{1}{K_{\nu}(x)}(\partial_x\Phi_{\nu}(x))^2 \}. \label{LL}
\end{split}
\end{equation}
In the above, $\nu=\rho,\sigma$ refers to charge and spin degrees of
freedom. We will drop this index from now on, as we will only be
concerned with the charge sector of the model.  We follow the
conventions used in Ref.~\onlinecite{seidel-05prb045113}, where the
field $\Phi(x)$
is related to the electronic density via
$\rho(x)=-\frac{2}{\pi}\partial_x\Phi(x)$, and $\Pi(x)$ is its
canonical momentum, $[\Phi(x),\Pi(x')]=i\delta(x-x')$.  The charge
part of the operator that creates a local right (upper sign) or left
(lower sign) moving electron is then proportional to
$\exp\{i\Theta(x)\pm i\Phi(x)\}$, where $\Theta(x)=-\frac{\pi}{2}
\int^x dx'\Pi(x')$, and the spin part is similar with charge fields
replaced by spin fields.

\begin{figure}[t]
  \begin{minipage}[c]{0.99\linewidth}
    \psfrag{B}{$\text{FL}$}
    \psfrag{A}{$\text{FL}$}
    \psfrag{C}{$\text{LL}$}
    \psfrag{D}{$K(x)$}
    \psfrag{E}{$K_L$}
    \psfrag{F}{$K_R$}
    \psfrag{L}{$\text{L}$}
    \psfrag{R}{$\text{R}$}
    \includegraphics[width=\linewidth]{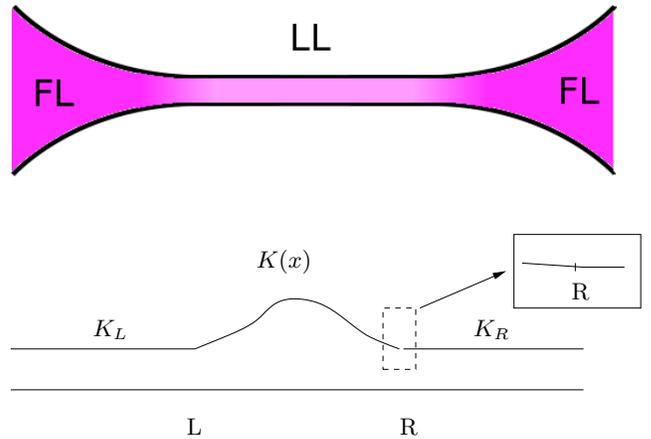}
  \end{minipage}
  \caption{(color online) Schematic plot of the Luttinger Liquid wire
    (LL) connected to two Fermi-liquid leads (FL). The value of $K$ is
    arbitrary inside the wire, but the left and right boundary values
    $K_L$ and $K_R$ are assumed equal to $1$ due to Fermi liquid
    screening effects. It is further assumed that these values are
    smoothly approached at the contacts (Eq.~\ref{bc0}).}
  \label{fig:twofermi}
\vspace{0pt}
\end{figure}

The Luttinger parameter $K(x)$ describes the electronic interactions,
where $K=1$ corresponds to the non-interacting case, and $K<1$ ($K>1$)
corresponds to repulsive (attractive) interactions. In the bulk of the
quantum wire, $K(x)$ and the mode velocity $v(x)$ may have arbitrary
spatial dependence. We will pay special attention to the case where
$K(x)$ smoothly approaches the non-interacting value $K=1$ at the
right and left contact ($x=x_L$ and $x=x_R$, respectively):
\begin{subequations}\label{bc0}
\begin{align}
  K(x_L)&= K(x_R)=1\label{bc0a}\\
  \partial_x K(x_L)&=\partial_x K(x_R) =0\label{bc0b}
\end{align}
\end{subequations}
This situation is depicted in Fig.~\ref{fig:twofermi}.  
The physical assumption behind these conditions is that the
modes of the
reflectionless wire extend into the leads,
where interactions are irrelevant\cite{maslov-95prb5539},
due to Fermi liquid screening.
Detailed arguments explaining why this picture applies to typical
experiments are also given in Ref.~\onlinecite{maslov04}.  Identical
or similar points of view have been adopted by many other authors
\cite{alekseev-98prl3503,lebedev-05prb075416,lehur-08ap3037}.  The
approach discussed here will confirm that while interactions may be
arbitrarily strong in the bulk of the wire, Eq. \eqref{bc0} is a key
ingredient to the observation of the quantized conductance. In
addition, one needs to impose a certain local chemical equilibrium
between the wire and the leads in the sense of Landauer, to be
discussed below.  Note that, independent of $K(x)$, $v(x)$ may
approach different limits at the left and right contact.  We also
point out that while in the infinite wire geometry discussed by other
authors~\cite{maslov-95prb5539,safi-95prb17040}, \Eq{bc0b} is
essentially {\em implied} (explicitly so in Ref.
\onlinecite{safi-95prb17040}) by the requirement that $K(x)$ tends to
a constant at $\pm\infty$, we find that the finite wire geometry makes
it necessary to impose \Eq{bc0b} as a separate constraint.  This
imposes stronger requirements on the strength of the proximity effect
than just \Eq{bc0a} alone.

\section{Equations of motion}
\label{sec:emo}
We will now proceed by working out the equations of motion for the
fields in~\eqref{LL}. From the equation of motion of $\Phi(x)$, one
obtains the continuity equation
\begin{equation}\label{cont}
  \partial_t \rho(x)= -\partial_x j(x)\;,
\end{equation}
where $j(x) =v(x) K(x) \Pi(x)$.  Furthermore, from the equation of
motion of $\Pi(x)$ we obtain
\begin{equation}\label{j1}
  \partial_t j(x) = -v(x)K(x)\;\partial_x\left(\frac{v(x)}{K(x)}\,\rho(x)\right).
\end{equation}
We shall now consider the effect of switching on a static electric
field by letting $ H\rightarrow H+\tilde{H}$, where :
\begin{equation}\label{hartree}
 \tilde{H}
=e \int dx\, \varphi(x) \rho(x)
=-\frac{2e}{\pi}\int dx\, \varphi(x) \partial_x\Phi(x).
\end{equation}
The underlying assumptions in writing down the model
Eqs.\eqref{LL}+\eqref{hartree} are the following: The potential
$\varphi(x)$ includes both the externally applied potential, as well
as the Hartree potential due to the response in the electronic charge
distribution.  Furthermore, we assume that the {\em difference}
between the true electron-electron interaction and the Hartree
mean-field term included in \Eq{hartree} can be represented by a local
contact interaction as usual in the long wavelength limit of a 1D
conductor.  This determines the position dependent Luttinger parameter
$K(x)$ discussed above.
We now take into account the effect of the potential $\varphi(x)$ on
the expression for the current, Eq.  \eqref{j1}. This equation then
takes on a form which is manifestly that of London's equation of a
superconductor:
\begin{equation}\label{london}
  \partial_t j = -\Lambda(x)^{-1}\;\partial_x\left(\mu(x)+e\varphi(x)\right),
\end{equation}
where $\mu(x)=\frac{\pi}{2}v(x)K(x)^{-1} \rho(x)$ can be identified as
the the local chemical potential in the wire in the absence of an
electric field, and $\Lambda(x)=\pi/(2v(x)K(x))$.  Eq.  \eqref{london}
makes it clear that in a steady state situation, any variation of the
electric potential in the wire has to be counter-balanced by the
chemical potential, in other words we have
\begin{equation}\label{const}
  \mu(x)+e\varphi(x) = \text{const.}
\end{equation}
in the wire. 
We emphasize that the quantity in Eq. \eqref{const}
is not directly related to the observed voltage
(which would then vanish).
Instead,
 $\mu(x)$
is just the average of the ``rightmoving'' and
``leftmoving'' chemical potential. 
In our ballistic wire, we do not
require that this {\em average} approaches the chemical potential of
the leads. Rather, we require that
\begin{equation}\label{bc}
  \mu_L=\mu^+(x_L)\;,\qquad \mu_R=\mu^-(x_R),
\end{equation}
where $\mu_L$ and $\mu_R$ are the chemical potentials in the left and
in the right lead, respectively, and $\mu^+$ and $\mu^-$ are the
``chemical potentials'' of the right- and leftmoving populations,
which we still need to define (Eq. \eqref{mupm} below).  
This situation is depicted in
Fig.~\ref{fig:fermielse}.
\begin{figure}[t]
  \begin{minipage}[c]{0.99\linewidth}
    \includegraphics[width=\linewidth]{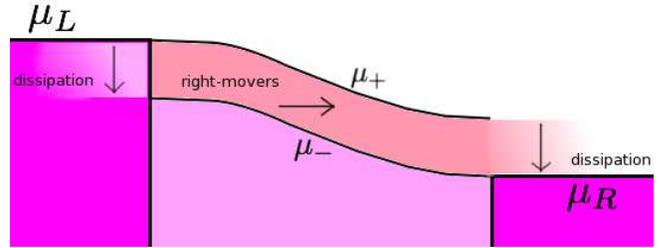}
  \end{minipage}
  \caption{(color online) Illustration of the Landauer-type boundary
    condition~\eqref{bc}. The right (left) moving eigenmodes are in
    chemical equilibrium with the electrons at the left (right) lead.}
  \label{fig:fermielse}
\vspace{0pt}
\end{figure}
The measured voltage is then the difference in the electro-chemical
potential between the leads:
\begin{equation}\label{V}
  V=\varphi_L+\mu_L/e-\varphi_R-\mu_R/e,
\end{equation}
where we have written $\varphi(x_{L,R})\equiv \varphi_{L,R}$.  This
quantity will be finite whenever a finite current is flowing.  Thus,
since no dissipation is assumed to occur within the wire, it is only
through a proper description of the contact between the wire and the
leads, where dissipation occurs, that the current-voltage relation can
be determined unambiguously
\cite{glazman-88jetp238,landauer87zpb,imry86}.  In order to proceed,
we now need to determine $\mu^+(x)$ and $\mu^-(x)$.  To this end, we
decouple the equations of motions, \ie \eqref{cont} and~\eqref{j1}.
(In this part of the calculation, the electric potential is not
important, so we will set it to zero.) We define~\cite{safi-95prb17040}:
\begin{equation}\label{rhopm}
  \rho_{\pm}(x)=\frac 1 2 (\rho(x)\pm j(x)/v(x)),
\end{equation}
which by~\eqref{cont} and~\eqref{j1} yields
\begin{eqnarray}\label{rhopm2}
  \partial_t\rho_\pm &=&\mp \partial_x( v(x)\rho_\pm(x) ) \nonumber \\ 
 && \pm \frac 1 2 (\partial_x
K(x) ) \frac{v(x)}{K(x)}(\rho_+(x)+\rho_-(x)).
\end{eqnarray}
We see that a complete decoupling is achieved only in parts of the
wire where $K(x)\approx \text{const}$ holds. In this case, we have a
complete decoupling between the quantities $\rho_-$ and $\rho_+$,
which can be interpreted as rightmoving and leftmoving densities,
respectively.  However, whenever we have a non-constant $K(x)$,
rightmovers will be scattered into leftmovers at some finite rate and
vice versa.  As explained above, we assume that $K(x)$ is
approximately constant (and equal to $1$) at the ends of the wire, as
a consequence of the proximity to the metallic leads.  It is only in
this case that we have a truly ``ballistic'' situation of independent
rightmoving and leftmoving populations with separate chemical
potentials, and that the boundary conditions \eqref{bc} at the
contacts are well defined in an obvious way.
\section{DC conductance}
\label{sec:dcc}
Independent of the spatial dependence of $K(x)$, however, we can write
the Hamiltonian \eqref{LL} (or rather, the charge part thereof) as a
sum of a $\rho_+$--part and a $\rho_-$--part:
\begin{equation}\label{Hpm}
  H=\frac \pi 2 \int dx\, \frac{v(x)}{K(x)} \left(\rho_+(x)^2+\rho_-(x)^2\right).
\end{equation}
Note that with the definition \eqref{rhopm}, Eq. \eqref{Hpm} holds
even where $K(x)$ is {\em not} constant, i.e. in regions of the wire
where the decoupling into left- and right-movers does not hold at the
equation of motion level.  Eq. \eqref{Hpm} naturally leads to the
definition of a local rightmoving and leftmoving chemical potential:
\begin{equation}\label{mupm}
  \mu^\pm(x)=\frac{\delta H}{\delta \rho_{\pm}(x)}=\pi
\frac{v(x)}{K(x)}\,\rho_{\pm}(x).
\end{equation}
As anticipated, we have $\mu(x)=\frac 1 2 (\mu^+(x)+\mu^-(x))$. We are
now in a position to address the main question: Are the boundary
conditions \eqref{bc} sufficient to uniquely specify the
voltage--current relationship? We start with \eqref{const} in the
following form:
\begin{equation}
  \mu(x_L)+e\varphi_L=\mu(x_R)+e\varphi_R.
  \label{acpot}
\end{equation}
Expressing $\mu$ through $\mu^+$ and $\mu^-$, we can bring this
equation into the form:
\begin{equation}
\begin{split}
&\mu^+(x_L)/e+\varphi_L-\mu^-(x_R)/e-\varphi_R=\\
&\frac 1 2 \left(\mu^+(x_L)-\mu^-(x_L)+\mu^+(x_R)-\mu^-(x_R)\right)/e.
\label{spplit}
\end{split}
\end{equation}
From~\eqref{bc} and~\eqref{V}, the left hand side is just the voltage
$V$ across the wire. As for the right hand side, we note that using
Eqs. \eqref{rhopm} and \eqref{mupm}, we have $j(x)=\pi^{-1}
K(x)(\mu^+(x)-\mu^-(x))$.  Given that $j(x)\equiv I/e$ is constant in
a steady situation, where $I$ is the electric current, the last
equation becomes
\begin{equation}\label{V2}
  V=\frac \pi {2e^2} (\frac{1}{K_L}+\frac{1}{K_R})\,I.
\end{equation}
Here, $K_{L,R}$ are the values of $K(x)$ at the contacts. From
different approaches, this expression has been obtained before in the
literature. As such, our framework also allows to describe generalized
reservoirs in~\eqref{V2} that are either of Fermi liquid or Luttinger
liquid type (encoded in $K_L$ and $K_R$), given that the proximity
boundary conditions are still justified. If we put back our original
assumption \Eq{bc0} that $K(x)$ approaches its ``Fermi liquid'' value
$1$ near the leads, i.e. $K_L=K_R=1$, we obtain the conductance
$G=I/V=\frac{e^2}{\pi}$. Or, putting back $\hbar$, which we had thus
far set equal to $1$:
\begin{equation}
\label{refgun}
  G=\frac{2 e^2}{2\pi\hbar}=2\frac{e^2}{h}.
\end{equation}
Hence, the conductance is quantized {\em independent} of the
interactions in the bulk of the wire, but {\em depending} on the
validity of the boundary conditions \eqref{bc0} and \eqref{bc} only.

An alternative derivation of this result using equations of motion has
been given by Safi in Ref. \onlinecite{safi97}, where the equations of
motion are argued to determine a particle transmission coefficient $T$. 
In our case, the use of
the boundary conditions \eqref{bc} allows us to completely avoid the
concept of single particle transmission. The same boundary conditions
are also used later by Safi in Ref.  \onlinecite{safi99ejp451}. Unlike
there, however, a complete decoupling of the dynamics of left- and
rightmoving degrees of freedom is not possible in the present context,
as evidenced by \Eq{rhopm2}.  Even earlier, \Eq{bc} has been employed
in Ref. \onlinecite{oreg-96prb14265} in an argument that is identical
to ours in the non-interacting case, but not otherwise.  We will
further comment on the relation of our work to
Ref. \onlinecite{oreg-96prb14265} below.

\section{AC conductance}
\label{sec:acc}

Aspects of the
dynamical response of Luttinger liquids and one-dimensional conductors in general have been frequently addressed
before~\cite{pretre}. In the time-reversal invariant case, the AC-conductance for
special interaction profiles has been studied by
Safi and Schulz\cite{safi-95prb17040}, and Safi\cite{safi97annal}.
Quantum
Hall
edges have been considered by Oreg and Finkel'stein\cite{yuval95}.
Here we derive a general result for the  time-reversal invariant case with arbitrary
interaction profile.
To this end, we will restrict
ourselves to frequencies much less than $v/L$, where $v$ is the
typical mode velocity.  Using Eq.~\ref{london}, in the presence of a
time-dependent current, Eq.~\ref{acpot} is modified according to
\begin{equation}
\mu(x_L) + e \varphi(x_L) - \mu(x_R) + e\varphi(x_R)= \int_{-L/2}^{L/2} dx\, \partial j_t(x) \Lambda(x).
\label{acmu}
\end{equation}
We assume the general mode expansion of the particle current
to be of the form
\begin{equation}
j_\omega(x,t) = e^{-i \omega t} \sum_k j_k(\omega) e^{ikx},
\label{mode}
\end{equation}
where the mode coefficients $j_k(\omega)$ are assumed to be analytic
functions of $\omega$. We know that in the DC limit, the current does
not depend on $x$, i.e.  $j_{k}(\omega)\vert_{\omega = 0} \propto
\delta_{k,0}$. This implies
\begin{equation}
 j_{k}(\omega) = \mathcal{O} (\omega)\;\;\; \forall k \neq 0.
 \label{finiteo}
 \end{equation} 
 With these assumptions, we plug Eq.~\ref{mode} into
 Eq.~\ref{acmu}. We keep only the leading term in $\omega$, which only
 depends on $j_{k=0} (\omega) \equiv \bar{I}(\omega)/e$. With this,
 Eq.~\ref{acmu} yields an AC generalization of Eq.\ref{spplit}:
\begin{eqnarray}\label{ACbalance}
&& \mu^+(x_L)/e + \varphi_L-\mu^-(x_R)/e -\varphi_R = -\frac{i \omega \pi \tau}{ e^2}\bar{I}(\omega) e^{-i\omega t}\nonumber \\
&&+\frac{1}{2 e}\left(   \mu^+(x_L) - \mu^-(x_L) + \mu^+(x_R) - \mu^-(x_R)\right), \nonumber \\
\end{eqnarray}
where the parameter $\tau$ is given by 
\begin{equation}
\tau=1/\pi \int_{-L/2}^{L/2} dx \Lambda(x),
\label{tau}
\end{equation}
with $\Lambda(x)$ as defined in~\eqref{london}.  As before, the
expression in the second line of Eq. \eqref{ACbalance} equals
$(I_R+I_L)\pi/2e^2$, where $I_R$ and $I_L$ denote the current in the
left and right lead, respectively. Due to possible capacitive effects,
we no longer assume $I_R = I_L$, but \emph{define}
$I(\omega)e^{-i\omega t} = (I_R(t)+ I_L(t))/2$.  From
Eq.~\ref{finiteo}, it is clear that $\bar{I}(\omega) = I(\omega) +
\mathcal{O}(\omega)$.  Identifying again the LHS of Eq.
\eqref{ACbalance} with $Ve^{-i\omega t}$, we finally find the
following relation between $V$ and $I(\omega)$:

\begin{equation}
I(\omega) = V \frac{2e^2 /h}{1-i\omega \tau}.
\label{AC}
\end{equation}
Strictly speaking, this relation is obtained only for the small
$\omega$ limit. However, we prefer to keep it in the above form which
is manifestly Drude's law. An interesting property of this low
frequency limit is that while non-universal corrections enter through
the parameter $\tau$ given by Eq. \eqref{tau}, details such as the
functional relation between the potential $\varphi(x)$ and the charge
distribution on the wire do not enter. These details would be
sensitive to aspects such as wire geometry, and could in principle
affect higher order corrections. In contrast, $\tau$ is solely
determined by the value of the Luttinger parameter within the wire.
This would allow one to compare $\tau$ as determined by AC
measurements to other means of extracting the Luttinger parameter.

\section{Discussion}
\label{sec:dis}
The derivation given here emphasizes that two ingredients are
sufficient to observe the quantized conductance in interacting quantum
wires. The ``proximity'' boundary condition \Eq{bc0} implies that the
interactions in the quantum wire become gradually weaker as the Fermi
liquid leads are approached, as originally emphasized in
Refs.~\onlinecite{maslov-95prb5539, safi-95prb17040}.  The
``chemical'' boundary conditions \Eq{bc} imply that the rightmoving
(leftmoving) electrons of the wire are in chemical equilibrium with
electrons of the left (right) reflectionless contact.  We note that if
we relax condition \eqref{bc0} by allowing $K(x_R)$ and $K(x_L)$ to
take on a common value $K\neq 1$ (and in particular if
$K(x)=K=\text{const.}$), \Eq{V2} yields the Kane-Fisher
result~\cite{kane-92prl1220}, according to which the conductance is
renormalized via $G=2Ke^2/h$.

We note that the physical assumptions leading to the quantized DC
conductance in our framework are compatible with those made by a
number of other workers (e.g. Refs.~\onlinecite{safi-95prb17040,
  maslov-95prb5539,
  alekseev-98prl3503,lehur-08ap3037,lebedev-05prb075416}). On the
other hand, alternative views are also found in the literature.
Notably, in Ref.~\onlinecite{oreg-96prb14265}, a quantized conductance
of $G=e^2/h$ is found for interacting spinless 1D fermions, within an
approach which does not directly involve boundary conditions of the
type \Eq{bc0}.  Instead, following Ref.
\onlinecite{kawabata95jpsj30}, it is argued
that the quantized value is always obtained when electron-electron
interaction effects are self-consistently included into the
definition of the total electric field. 
We will now clarify the relation of
this approach to ours (see Ref. \onlinecite{safi97} for a different discussion).
 We first observe that such interaction effects
also enter the definition of our total voltage, \Eq{V}.  Here, the
Hartree mean field part of the interactions is included in the
potential $\varphi(x)$, while correlation effects are included in the
definition of the right- and left-moving local chemical potential,
\Eq{mupm}. In addition, however, these chemical potentials do not
depend on interactions alone, but equally depend on the kinetic part
of the Hamiltonian, through their dependence on the local density.
This is in accordance with the general notion that in gapless 1D
systems, kinetic energy and interactions should be treated on equal
footings, as is manifest by their form in the Luttinger Hamiltonian.

The difference between the approach discussed here and those
emphasizing self-consistent fields over boundary conditions of the
form \Eq{bc0} can be understood as follows.  It is useful to discuss
both spinless and spinful fermions for a {\em homogeneous} wire with
$K(x)=K=const$.
In the spinful case, our result reduces to that of Kane and Fisher,
$G=2Ke^2/h$.  This value coincides with that for non-interacting
spinless fermions, $e^2/h$, if $K$ equals $1/2$.  This again is
entirely consistent with the well-known fact that for $K=1/2$, the
charge part of \Eq{LL} is identical to the Hamiltonian of
non-interacting spinless fermions. (This remains true in the presence
of an electromagnetic field.)  Thus, in the absence of boundary
conditions, at least naively nothing should distinguish the two cases.  If, on the
other hand, one concludes that the conductance is unrenormalized, even
in the homogeneous wire case with unspecified boundary conditions, one
would obtain $G=2e^2/h$ for spinful fermions even at $K=1/2$, whereas
$G=e^2/h$ for non-interacting spinless fermions.  
This may  seem somewhat
at odds with the fact that the charge part of the Hamiltonian is
exactly the same in both cases.  
The only difference in these two
cases is the decomposition of the Hamiltonian into distinct kinetic
and interacting parts.  In the present approach, where these parts
are treated on equal footing throughout, a difference in conductance
may only arise through different boundary conditions.
Indeed, with our conventions, these boundary conditions must read
$K_L=K_R=1/2$ if the system, including leads, consists of spinless
fermions. 
On the other hand, a major difference between our treatment
and that in Refs. \onlinecite{kawabata95jpsj30, oreg-96prb14265}
lies in the definition of the measured voltage.
While we include a chemical potential drop between the leads,
which is related to a chemical potential difference between rightmovers
and leftmovers by \Eq{bc}, in Refs. \onlinecite{kawabata95jpsj30, oreg-96prb14265}
 the voltage
is just the integral over the total electric field.
This may be a more natural definition 
in a truly homogeneous wire without boundaries.
The chemical potential difference included into our voltage
is always finite (for the homogeneous wire) in the presence of a finite current.
This explains why in the homogeneous wire case, our result for the conductance
(renormalized) will in general differ from that of Refs. \onlinecite{kawabata95jpsj30, oreg-96prb14265} (unrenormalized).
We do believe that in most experiments conducted so far, where a voltage is
measured between two leads and Landauers basic ideas apply, 
the voltage definition used here is the appropriate one, and
the observed conductance is thus entirely determined
by boundary conditions of the kind used here.
We note, however, that experimental setups have been proposed~\cite{maslov04}
which circumvent the application of leads altogether by measuring
alternating currents induced in the wire by a resonator.
The approach of Refs. \onlinecite{kawabata95jpsj30, oreg-96prb14265}
may indeed be better suited to describe this case.
Such
experiments would hence be very useful to distinguish the picture developed
here and elsewhere from others.

We finally remark that although we did not develop a detailed scenario
for possible deviations from the quantized value of the conductance
\cite{matveev04prb245319,danneau-08prl016403,thomas-96prl135}, we are
confident that the explanation for such deviations may be addressed
similarly if proper boundary conditions describing these cases are
identified.  
One interesting possibility is that a spin gap exists in the entire wire including the contacts.
In this case one may
conjecture that the spinless case effectively applies with an
effective charge $e^\ast=2e$ (Ref. \onlinecite{KYpriv}, c.f. also
Ref.~\onlinecite{seidel-05prb045113}).  However, the interactions can
then not be considered weak at the boundary, and we defer a detailed
analysis to further studies.
Moreover, we conjecture that our treatment of the AC
conductance largely carries over to these cases. Since deviations from
the $\omega\rightarrow 0$ limit are expressible solely through bulk
properties of the wire, the natural modification of Eq. \eqref{AC}
would be to simply replace the numerator by the respective
non-universal value of the DC conductance.  The Drude-type expression
\eqref{AC} we find in the AC case may also suggest that this
expression could be obtained in an "RPA-like" approximation. If so,
this may suggest that the validity of this expression can be extended
beyond the $\omega\ll v/L$ regime.  We note, however, that $v/L$
typically well exceeds the gigahertz range under relevant conditions,
and so the restriction $\omega\ll v/L$ should not pose a stringent
limitation for current experiments.



\section{Conclusion}
In this article we addressed the fundamental nature of quantized
conductance in one-dimensional quantum wires.
Working in a framework based on the analysis of the equations of
motion of an inhomogeneous interacting Luttinger liquid, we argued for
the significance of two types of boundary conditions as the underlying
cause for a quantized conductance.  The formalism presented here
simplifies some earlier treatments without loss of essential physics, and
should serve as a useful starting point also when discussing possible
deviations from the ideal quantized conductance.  We have further
studied AC deviations from the universal conductance, and found that
these are expressible through general Luttinger parameters in a simple
manner in the low frequency limit.

\begin{acknowledgments}
  We are indebted to D.-H. Lee and Y.-C. Kao for bringing this problem
  to our attention.
  RT acknowledges insightful discussions with D.~B.~Gutman, as well as
  I.~Affleck and C.-Y.~Hou at the LXXXIX Les Houches Summer School.
  AS would like to thank K.  Yang  for stimulating
  discussions, and M. Grayson for insightful comments and a careful reading of an earlier version of this
  manuscript. We also thank I. Safi for very helpful remarks.  RT is supported by a Feodor Lynen fellowship of the
  Humboldt Foundation and by the Sloan Foundation.  AS is supported by
  the National Science Foundation under NSF Grant No. DMR-0907793.
\end{acknowledgments}


\end{document}